\documentstyle[preprint,revtex]{aps}
\begin{document}
\preprint{IMSc 92/38}
\preprint{cond-mat@babbage/9209017}
\begin{title}
Numerical Study of the Wheatley-Hsu-Anderson Interlayer\\
Tunneling mechanism of High $T_c$ Superconductivity
\end{title}
\author{Mihir Arjunwadkar}
\begin{instit}
Department of Physics, University of Poona\\
Pune 411 007 INDIA
\end{instit}
\author{G. Baskaran, Rahul Basu and V. N. Muthukumar\cite{email}}
\begin{instit}
The Institute of Mathematical Sciences, Madras 600 113 INDIA
\end{instit}
\begin{abstract}
We present results obtained (by exact diagonalization) for the problem of
two $t-J$ planes with an interlayer coupling $t_\perp$. Our results
for small hole concentrations show that in-plane superconducting
correlations are enhanced by $t_\perp$. When the constraint on
double occupancy in the $t-J$model is relaxed, the enhancement disappears.
These results illustrate the inter--layer tunneling
mechanism for superconductivity.
\end{abstract}
\newpage
Ever since Anderson's  proposal \cite{pwa} that the physics of the high $T_c$
superconductors is contained in the one band Hubbard Hamiltonian
\begin{equation}
H= -t\sum_{i,j} c^\dagger_{i\sigma}c_{j\sigma}+
   U\sum_i n_{i\uparrow}n_{i\downarrow}
\end{equation}
in the limit of large $U$, the Hubbard model and its derivative, the
$t-J$ model
\begin{equation}
H = -t\sum_{<i,j>}c^\dagger_{i\sigma}(1-n_{i-\sigma})c_{j\sigma}
    (1-n_{j-\sigma})+J\sum_{<i,j>}(\vec S_i.\vec S_j-{1\over 4}n_in_j)
\end{equation}
where $J = \frac{4t^2}{U}$\\
have been the subject of several analytical studies. Despite these
attempts, a complete quantitative understanding of the ground state of these
Hamiltonians is still lacking. Finite size clusters of these models have
also been analyzed extensively \cite{dag} by exact diagonalization and
Variational Monte Carlo techniques. These studies indicate that several
of the anomalous normal state properties of the $CuO$ superconductors
could be accounted for by the $t-J$ Hamiltonian. However in these
numerical studies, there are no robust signals of a superconducting
phase in either the Hubbard or the $t-J$ model \cite{hirsch}.
In our view this is hardly
surprising since superconductivity in the $CuO$ compounds is governed by
a scale that is neither $t$ nor $J$, but the interlayer coupling
$t_\perp$. This was first proposed by Wheatley, Hsu and Anderson
\cite{wha}(WHA).

We now describe the basic physics behind the WHA mechanism . There is now
increasing evidence that the normal state of the high $T_c$
superconductors is a non-Fermi liquid\cite{pwa1}. The low energy excitations
are spin--charge decoupled (the spinons and holons) i.e. real
electron like quasi-particles having both spin and charge are absent
asymptotically at the Fermi energy. This phenomenon is called
``confinement'' of real electrons close to the Fermi surface.
These features are known to be present in the 1--d Hubbard model as has
been discussed, for example, by Anderson and Ren \cite{andren}.
It is believed that the 2--d  Hubbard model or the
$t-J$ Hamiltonian would reproduce these features. Because of
spin--charge decoupling or the confinement of electron like
excitations, single electron motion between two $CuO$ planes
(which necessitates real scattering processes between the spin and
charge degrees of freedom) gets suppressed i.e., the first order
process in $t_\perp$, which transfers a real electron from a plane to a
neighboring plane, becomes ineffective asymptotically at the fermi
level \cite{pwa2}. However, the
second order processes generated by $t_\perp$ are not suppressed. This
is because the $t_\perp$ term to second order gives rise to processes
that do not leave an unpaired spin in a layer but creates/annihilates a
spinon pair comprising an ``up'' and a ``down'' spinon in a singlet
state. Since in the RVB
ground state such spinon pair fluctuations are present, these processes
are not suppressed.  These can be seen within the framework of second
order perturbation theory \cite{vnm} which generates a term of the form
$c^{l\dagger}_{i\uparrow}c^{l\dagger}_{j\downarrow}c^m_{j\downarrow}
c^m_{i\uparrow} + h.c.$, which can be rewritten
as $b^{l\dagger}_{ij}b^m_{ij}$ where $b_{ij}$ are the usual singlet
operators (here $l$ and $m$ label different layers).

It therefore follows that the
processes that are not suppressed are those that transport a pair of
electrons in a singlet state between the $CuO$ planes. Since these are
precisely the superconducting fluctuations, we conclude that the effect
of the $t_\perp$ is to cause electron pair tunneling between the layers
which in turn causes superconductivity.  In the language of
spinons and holons, this implies that the $CuO$ plane already supports
spinon pair excitations as the spinons form a paired condensate.  The
inter plane coupling causes pairs of
holons to hop between layers by creating or annihilating spinon pairs.
i.e. there is a ``leakage'' of ODLRO to the holons from the spinons
caused by the inter plane coupling $t_\perp$ \cite{vnm}.
 Insofar as {\em
spin quantum numbers} are concerned, a pair of electrons (holes) in a
singlet state is the same as a pair of antiholons (holons). Thus
electron (hole) tunneling to second order causes antiholon (holon)
pair tunneling.

{}From these arguments it is clear that the occurence of superconductivity
is crucially related to two factors:
\begin{enumerate}
\item[(i).]
a large $U$ (or irrelevance of double occupancy)
which causes spin-charge decoupling and
\item[(ii).]
the effect of $t_\perp$ on the the $2-d$ planes having spin-charge
decoupling.
\end{enumerate}
Since our main results concern (ii), let us briefly consider relaxing
the requirement (i). Then the problem reduces to that of two coupled
$t-J$ planes with no constraint on double occupancy. The problem is
also similar to that of two conventional BCS-superconducting
planes coupled by $t_\perp$.
Since $U=0$, there is no suppression of
single electron motion between the planes. This causes $t_\perp$
to act as a ``pair breaker'' i.e. the pairing energy can be lost at the
expense of kinetic energy gained by hopping between the planes. This
is the reason that mean field studies (where the single--occupancy
constraint is not imposed exactly) on coupled $t-J$ planes show a
decrease in $T_c$ with $t_\perp$\cite{shanks}. In our numerical study,
we find that
$t_\perp$ causes superconducting correlations to decrease marginally.
On the contrary, in the case of $t-J$ planes
with no double occupancy, we see that the
inter--plane coupling {\it enhances} in--plane superconducting correlations.
We therefore conclude that $t_\perp$ can enhance superconducting
correlations only when the condition $n_{i}=0\ {\rm or}\ 1$ is imposed
exactly.

We have performed exact diagonalization \cite{call} studies on $4+4$,
$5+5$ and $6+6$ site clusters. These clusters consist of two planes
coupled by the $t_\perp$ term. We use periodic
boundary conditions. In addition we use two different geometries for the
$6+6$ case :(a) for a closed chain and (b) a grid. We have presented
results only for the grid geometry since the results for the closed
chain are qualitatively similar.  We have chosen, on a scale
of $\vert t\vert = 1$, $J=0.31$, and varied $\vert t_\perp\vert$
from $0$ to $0.9$.  Our results are therefore of direct relevance to the
$CuO$ compounds in the region of small $t_\perp$. To illustrate the
effect of $t_\perp$ in the absence of $U$, we have also diagonalized a
$4+4$ cluster with two holes after relaxing the constraint on double
occupancy. We give below a short
description of the numerical aspects of the computations.

We have used a basis for the Hamiltonian matrix in which $S_z$ is
diagonal, so that the only diagonal matrix elements are those of the
$\sum_{<i,j>}S^z_iS^z_j$ term. The off-diagonal elements come from the
$\sum_{<i,j>}S^+_iS^-_j + S^-_iS^+_j$ and the $t$ and
$t_\perp$ terms. Note that the off--diagonal elements
are mutually exclusive in the sense that a given pair of distinct basis
states can at best be connected by one of these three off--diagonal terms.
In addition, this basis also excludes all states with double occupancies.
We compute the lowest eigenvector if the ground state is
non-degenerate and all the degenerate ones if the ground state is
degenerate (which it typically is for $t_\perp=0$)using the conjugate
gradient method \cite{cg}. We use a simulated annealing based algorithm
\cite{md} to get an improved starting guess for the conjugate gradient.

To look for superconductivity we compute the extended--singlet correlation
function as defined by Hirsch \cite{hirsch} which we explain below.\\
Let
$$
b_{ij} = {1\over \sqrt{2}}[c_{i\uparrow}c_{j\downarrow}
                         - c_{i\downarrow}c_{j\uparrow}],
$$
where $(i,j)$ are nearest neighbor sites in a plane. Then the
extended--singlet pairing correlation (SPX) is defined as
\begin{equation}
\chi = \frac{1}{N}\sum_{<i,j><k,l>}\langle b_{ij}b^\dagger_{kl}\rangle
\end{equation}
where $\langle\cdots\rangle$ represents expectation value in the ground
state. Here $N$ is the number of in--plane sites. If $\chi$ scales
as $N$, then the result suggests a superconducting
instability in the thermodynamic limit. (We look only for s-wave
pairing).

The results of the above computations show several interesting features
that are size--independent (see Fig. 1,2 and 3). First, we note that for
the case of two holes, the SPX always increases with $t_\perp$. For
cluster sizes $4+4$, $5+5$ and $6+6$, this corresponds to a doping of
$\simeq$ 25\%, 20\% and 16\% respectively. As soon as we add two more holes,
we enter a region of large doping. The results with four holes therefore
show a qualitatively different behavior. The SPX in this case is not
affected much by $t_\perp$. The results resemble those of the
unconstrained $t-J$ model. The same behavior persists for larger hole
concentrations.
In this sense, we suggest that we have
crossed over from a non--Fermi liquid phase (which is sensitive to
$t_\perp$) to a phase which is less of a non--Fermi liquid.

Next we consider the limit $t_\perp \rightarrow 0$. In this limit with
two holes in the system, the ground state has one hole in each layer on
an average. Therefore the contribution to SPX is dominated by terms of
the form $<O_{12}O_{12}^\dagger >$, $<O_{12}O_{23}^\dagger>$.
Such essentially on--site correlations are not related to superconducting
order in the thermodynamic limit. In previous studies \cite{dag} of the
$t-J$ model, it was noticed that only such terms contributed to the SPX.
This led to the conclusion there are no incipient long--range (superconducting)
correlations in the $t-J$ model. Our results for $t_\perp = 0$ also
reflect this. (Also note that for the case of $4+4$ with 2 holes and
with no constraint on double occupancy, the SPX is maximum for
$t_\perp=0$. As discussed earlier, this shows the ``pair breaking''
nature of $t_\perp$.)

However as $t_\perp$ increases, the long--range correlations increase
rapidly. In fact, it is this behavior of the long--range correlations that
causes the enhancement of the total SPX with $t_\perp$. To show this, we
have subtracted in the expression for SPX, those terms with none of the
indices $<i,j>, <k,l>$ in eq.(3) equal i.e. terms wherein bonds $<i,j>$ and
$<k,l>$ neither overlap nor touch, and examined the resulting behavior of
SPX  with $t_\perp$ for $6+6$ with two holes. The results are shown in Fig. 4.
The figure demonstrates the dramatic increase
of the ``long range'' part of the pair susceptibility by a factor of 30. A
similar increase is noticed for the $4+4$ and $5+5$ cases.
We observe similar results for a large $U$ Hubbard model
where the enhancement of pair susceptibility due to interlayer tunneling
is clearly visible.

Finally we address the question of finite size scaling. Comparison of
our data for the $4+4$ and $6+6$ cases shows a scaling which is
slightly smaller than $N$. From our results it is also clear that $5+5$
is quantitatively different. This we believe to be an even--odd feature.

Since we have not diagonalized larger clusters, we are aware
that these results are not conclusive with regard to scaling.
However we believe that these preliminary results are very encouraging
and point in the right direction towards further numerical studies. For
it is clear from the definition of the SPX and our criterion for
superconductivity that any instability in the thermodynamic limit has to
come from these long range correlations. In our study, we find that the
contribution to the SPX (for a given $t_\perp  \neq 0$)
from the long range correlations increases with
size. This suggests the presence of a superconducting phase in the
thermodynamic limit.
Another limitation we have faced is in varying the number of holes. The
crossover from small doping to the overdoped case needs more careful
scrutiny. But again our results suggest that the feature indeed
exists.

One can also question whether the small systems
we have investigated can exhibit the physics of spin--charge decoupling.
It is difficult to answer this question quantitatively. However the recent
numerical work by Jagla and co--workers \cite{jagla} clearly demonstrates
the phenomenon of spin--charge decoupling in a finite size system.

To summarize, we have addressed the question of the effect of the
interlayer coupling $t_\perp$ on in--plane superconducting correlations
(SPX) in the context of the $t-J$ model with the  constraint on double
occupancy imposed exactly. Our results for two holes in $4+4$,
$5+5$ and $6+6$ sites show that these correlations are increased by
$t_\perp$. This happens because of the rapid increase of the long--range
correlations with $t_\perp$. This feature disappears as we increase the
number of holes or when the constraint on double occupancy is relaxed.
The former suggests the existence of a crossover
point between small and heavy doping regimes based on this contrasting
behavior of SPX as a function of $t_\perp$.
Work is currently in progress in evaluating
dynamic correlation functions and the effect of a magnetic field on
these correlations.

\acknowledgements
We thank D. G. Kanhere for several useful suggestions. One of us (M. A.)
would like to acknowledge hospitality at
the Institute of Mathematical Sciences where part of
this work was done. Most of the computations were done on a Sun
Sparcstation 1 (DST project SBR 32 of the National Superconductivity
Programme) and an HP9000/835 machine (DST project SP/S2/K22/87).
\eject

\eject

{\Large \bf Figure Captions}
\begin{enumerate}
\item
SPX for the $4+4$ cluster. The dashed line represents the behavior of
SPX when the constraint on double occupancy in the $t-J$ model is relaxed.
\item
SPX for the $5+5$ cluster.
\item
SPX for the $6+6$ cluster.
\item
Long range correlations in  $6+6$ cluster with two holes
(see text for explanations).
\end{enumerate}
\end{document}